\documentclass{article}

\usepackage{arxiv}

\usepackage[utf8]{inputenc} 
\usepackage[T1]{fontenc}    
\usepackage{hyperref}       
\usepackage{url}            
\usepackage{booktabs}       
\usepackage{amsfonts}       
\usepackage{nicefrac}       
\usepackage{microtype}      
\usepackage{lipsum}
\usepackage{graphicx}
\usepackage{amsmath}
\usepackage{subfig}

\usepackage{float}
\usepackage{hyperref}
\usepackage{siunitx}
\usepackage{amssymb}
\usepackage{cleveref}
\usepackage[inline]{enumitem}
\usepackage{multirow}
\usepackage[table,xcdraw]{xcolor}

\newcommand{\BGpre}[1]{\widehat{\mathrm{bg}}_{#1}(t)}
\newcommand{\BGpreAvg}[1]{\overline{\widehat{\mathrm{bg}}}_{#1}}

\newcommand{\BGact}[1]{\mathrm{bg}_{#1}(t)}
\newcommand{\BGactAvg}[1]{\overline{\mathrm{bg}}_{#1}}

\DeclareSIUnit\BGunit{\milli\gram\per\deci\liter}

\newcommand{\etal}{\textit{et al.}\ }

\usepackage{xcolor}
\begin{document}

\title{A Critical Review of the state-of-the-art on Deep Neural Networks for Blood Glucose Prediction in Patients with Diabetes}
\author{F\'elix Tena, Oscar Garnica, Juan Lanchares, J. Ignacio Hidalgo
\thanks{This work was supported by Fundaci\'on Eugenio Rodr\'iguez Pascual 2019 grant --Development of Adaptive and Bioinspired Systems for Glycaemic Control with Continuous Subcutaneous Insulin Infusions and Continuous Glucose Monitors; the Spanish Ministerio de Innovaci\'on, Ciencia y Universidad --grant RTI2018-095180-B-I00;
Madrid Regional Government --FEDER grants B2017/BMD3773 (GenObIA-CM) and Y2018/NMT- 4668 (Micro-Stress- MAP-CM).}
\thanks{F. Tena (feltena@ucm.es), O. Garnica (ogarnica@ucm.es), J. Lanchares (julandan@ucm.es) and J.I. Hidalgo (hidalgo@ucm.es) are with the Dpt. of Computer Architecture, Facultad de Informática, Universidad Complutense de Madrid. J.I. Hidalgo is also with the Instituto de Tecnolog\'ia del Conocimiento.}}


\maketitle

\begin{abstract}
This article compares ten recently proposed neural networks and proposes two ensemble neural network-based models for blood glucose prediction. All of them are tested under the same dataset, preprocessing workflow, and tools using the OhioT1DM Dataset at three different prediction horizons: 30, 60, and 120 minutes. We compare their performance using the most common metrics in blood glucose prediction and rank the best-performing ones using three methods devised for the statistical comparison of the performance of multiple algorithms: scmamp, model confidence set, and superior predictive ability. Our analysis highlights those models with the highest probability of being the best predictors, estimates the increase in error of the models that perform more poorly with respect to the best ones, and provides a guide for their use in clinical practice.
\end{abstract}

\section{Introduction}

Diabetes mellitus (DM) or simply \textit{Diabetes} is a group of metabolic disorders of multiple aetiology characterized by the presence of high concentrations of glucose in blood (BG), a.k.a. hyperglycemia. It comes with disturbances of carbohydrate (CH), protein, and fat metabolism resulting from defects in insulin secretion, insulin action, or both \cite{ghosh}. 

Patients with diabetes  are classified into two main groups depending on the anomalies that cause high BG levels: insufficient insulin production, Type 1 diabetes (T1DM) or insulin resistance, Type 2 (T2DM) \cite{idf}. We focus on patients with T1DM, who  need to compensate for the absence of insulin secretion by administering exogenous artificial insulin. If the amount of insulin administered is not enough to process the ingestion of food, glucose levels will remain at high values. If this situation is maintained for a long period of time, multiple complications may appear. Acute hyperglycemia symptoms such as frequent urination, thirst, headache, or fatigue among others,are related to dehydration as the kidneys try to filter  excess glucose. If hyperglycemia is not treated it can produce ketoacidosis, which produces weakness, confusion, or  even coma \cite{mayoHyper}.

On the other hand, an insufficient ingestion of CH in relation with insulin administration leads to hypoglycemic events. As BG levels decrease, autonomic nervous system activity increases, thus initiating warning signs such as anxiety, sweating, hunger or palpitations \cite{gerich}. If it is not treated it can produce muscle weakness, inability to drink or eat, convulsions, unconsciousness, or even death \cite{mayoHypo}.

The effects of the disease can be countered with a healthy lifestyle, continuous glucose monitoring, and close follow-up actions, which as an outcome promotes the patient's well-being and reduces medical costs \cite{hidalgo}. This control is a challenging task for the patient since she/he somehow needs to substitute the action of a healthy pancreas. Patients have to determine BG levels at different times of the day, monitor CH intakes, and administer the appropriate insulin doses (fast insulin bolus injections to cover the CH ingested and slow-action insulin administration to cover the basal production of insulin). The objective is to maintain healthy glucose levels, from \SIrange{70}{180}{\BGunit} \cite{gerich}. Continuous glucose monitoring systems (CGM) monitor interstitial glucose levels every five to fifteen minutes automatically \cite{meijner}, and in combination with insulin pumps facilitate the control of BG levels. 

Nevertheless, the prevention of hyperglycemic and hypoglycemic events requires BG values to be forecast ahead of time because of the lag in the effects of corrective actions. However, dynamic predictive BG  models are difficult to develop because of the lack of a general response of the body to the different variables which are affected by the particularities of each patient. Classic glucose models use linear equations and define profiles that do not cover these particularities \cite{hidalgo}. As aforementioned, CGMs provide glucose time series, which can be analyzed using time series techniques. The first two prediction systems with enough accuracy to be implemented in clinical treatments for patients with T1DM \cite{wiley} are the Support Vector Regression (SVR) and AutoRegressive Integrated Moving Average (ARIMA) models, which are usually applied as a base case to compare the performance of new BG forecasting models \cite{bunescu2013}.

More generally, these CGM series can be used as input datasets for BG accuracy prediction using machine learning techniques. Hence, during recent years, many machine learning techniques have been applied to BG prediction, attempting to obtain better results than the previously mentioned models, such as those carried out in the past by our team (Adaptive and Bioinspired Systems Research Group), which are based on grammatical evolution and which have also produced good results \cite{hidalgo2017data}. Among them, the most promising ones are those based on neural networks (NN). However, it is difficult to extract feasible and judicious conclusions from all of these studies because of a lack of common patient datasets, data pre-processing, missing handle policies, sample rates, feature engineering, forecasting horizons, and equivalent metrics. We tackle the development of this meta-study to compare the performance of the most recent NN proposals for BG prediction using a common framework for the comparison (using the same datasets, the same features, the same sampling rates, and the same metrics) that enables the judicious use of the current best evidence in making the best decisions about patient care. Hence, we study how current state-of-the-art NN models behave for BG predictions to identify which one is most suitable, if any, for each of the possible scenarios present in the datasets.

This paper is structured as follows. \Cref{sec:nn} is a brief introduction to neural networks and \Cref{sec:models} presents the NN models we compare in this work. Next, \Cref{sec:experimental_results} is devoted to presenting our feature engineering with the OhioT1DM Dataset and the experimental results. Finally, \Cref{sec:conclusions} summarizes the conclusions and findings of this work.

\section{Artificial Neural Networks} \label{sec:nn}

A neural network (NN) is an interconnected assembly of simple processing elements, called units or neurons, in a structure that mimics the organization of the neurons in human brains \cite{gurney}. Its purpose is to process information with a series of mathematical operations, with dynamic responses, to recognize patterns in data that are too complex to be manually identified by humans. Specific NN structures are more suitable to each type of problem, so the first step to using an NN is to define its structure; that is, to define the number and type of layers. A layer is a set of neurons such that the neurons' output in one layer is the neurons' input is the next layer. Three main kinds of layers are defined \cite{nielsen}:
\begin{itemize}
    \item The input layer is the first layer of neurons and receives the NN's external input. It sorts the information to be processed by the NN, and, for this reason, its structure (type and number of neurons) is determined by the dataset's features.
    \item The hidden layers are a group of layers in which all the transformations are done. All the input and output are internal variables of the NN; they are not visible outside of the NN.
    \item The output layer is the last one. It returns the NN's outcome.
\end{itemize}

Different types of neurons have been proposed. Typically, all neurons in a single layer have the same type, although different layers can have different neuron types. The top three most common types of neurons lead to the different architectures of NNs, which are described below.

\begin{figure}[htb]
    \centering
    \includegraphics[width = 2.4in]{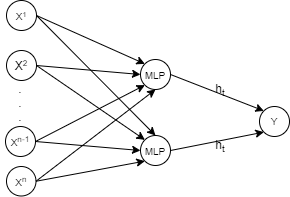}
    \caption{MLP structure.}
    \label{fig:mlp_arch}
\end{figure}

Deep feedforward networks, a.k.a. multilayer perceptrons (MLPs) or dense NNs, are the most well known since they are the basis on which neural networks are studied, and they are widely used for classification problems. In MLPs, data flow from the input to the output straightforwardly without any internal loop \cite{goodfellow}, as illustrated in \Cref{fig:mlp_arch}.

The $i$-th neuron receives an input vector at time $t$, $\mathbf{x}_t^i$, and outputs a value, $h_t^i$ calculated using a function called the activation function, $g(\cdot)$. The three most common activation functions are ReLU, sigmoid, and hyperbolic tangent, \Crefrange{eq:relu}{eq:tanh}, respectively. In this paper, we also use linear and exponential linear units, \Crefrange{eq:linear}{eq:ELU}.

\begin{align} 
  g(x) \equiv \mathrm{ReLU}(x) &= \max\{0,x\} \label{eq:relu} \\ 
  g(x) \equiv \sigma{(x)} &= \frac{1}{1+\exp(-x)} \label{eq:sigmoid} \\
  g(x) \equiv \tanh{(x)} &= \frac{\exp{(2x)}-1}{\exp(2x)+1} \label{eq:tanh} \\
  g(x) &= x \label{eq:linear} \\
  g(x)\equiv \mathrm{ELU}(x) &= 
      \begin{cases} 
       x                       & , x > 0 \\
       \alpha \cdot (e^{x} -1) & , x \leq 0
      \end{cases} \label{eq:ELU}
\end{align}

\Cref{eq:afin} presents the affine transformation in an MLP neuron. $\mathbf{x}_t^i$ is multiplied by an internal neuron parameter called the weight vector, $W^i$, and shifted by a vector called bias, $b^i$. The weights and biases are the dynamic elements modified in the training phase to adjust the output with respect to the input data and make the corresponding prediction.

\begin{equation}
     h_t^i = g(\mathbf{x}_t^i \cdot W^i + b^i)
     \label{eq:afin}
\end{equation}

\begin{figure}[H]
    \centering
    \includegraphics[width = 2.8in]{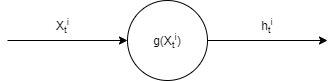}
    \caption{MLP neuron.}
    \label{fig:mlp}
\end{figure}

Convolutional Neural Networks (CNN) are specialized NNs to process grid-like data. CNNs initially targeted image recognition \cite{goodfellow}, and recently they have also been used in time series analyses \cite{brunel}. 

\begin{figure}[H]
    \centering
    \includegraphics[width = 2.8in]{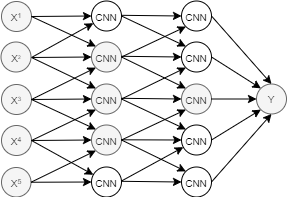}
    \caption{CNN structure.}
    \label{fig:cnn_arch}
\end{figure}

CNNs use the same neuron type and activation functions as MLPs. What differs from MLPs is, firstly, that CNNs use the convolution operation on $\mathbf{x}_t^i$ instead of affine transformations; that is,

\begin{equation}
     h_t^i = g(\mathbf{x}_t^i \circledast w^i + b^i)
     \label{eq:conv}
\end{equation}

where $w$ is called the filter, and it is shifted by the bias vector. Secondly, it differs in the relation among neurons in different layers. \Cref{fig:cnn_arch} presents the structure of a CNN and illustrates how a CNN neuron is not connected to all the output from the previous layer nor is its output connected to all the input in the next one, this being more efficient as the last layer indirectly receives information from all the input.

\begin{figure}[H]
    \centering
    \includegraphics[width = 2.4in]{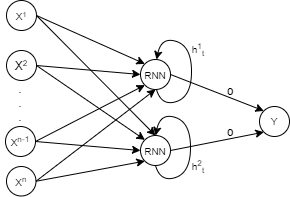}
    \caption{RNN architecture.}
    \label{fig:rnn}
\end{figure}

The third NN architecture is the Recurrent Neural Network (RNN). Unlike MLPs and CNNs, data do not flow linearly through the RNN since RNNs have loops within their neurons, so some of the previous neurons' output is used as input in the next timestep. \Cref{fig:rnn} illustrates the typical RNN architecture, and \Cref{eq:rnn} describes the input vector transformation within an RNN cell,
\begin{equation}
         h_t^i = g(\mathbf{x}_t^i \cdot w^i + h^i_{t-1} \cdot w^i_{h,t} + b^i)
     \label{eq:rnn}
\end{equation}

where $h_t^i$ is the hidden state of the $i$-th neuron; this is short-term memory because this value is stored for one iteration; it is multiplied by $w_{h,t}$ which is the weight for the hidden state. In \Cref{fig:rnn}, $o$ is the output state vector, which is the information passed to the next layer, which can be a vector built by all the hidden states of all the iterations or only the last iteration, according to the needs of the system.

RNNs aim to keep an internal state that summarizes the input history, so the current NN's output also depends on the previous NN's input. This architecture is especially suitable for problems in which the actual output depends on previous input history, such as text recognition, text prediction, or short-term or long-term dependence time series forecasting. Critical Vanilla RNN problems are related to the ``flowing backward in time" problems that appear with long-term dependence time series. Long-Short Term Memory (LSTM) NNs are a kind of RNN that solves the backpropagation vanishing or exploding gradient by enforcing constant error flow \cite{hochreiter}. Hence, while conventional RNNs can stand up to 10 discrete time steps without vanishing or exploding, LSTM can operate with more than 1000 timesteps \cite{gers}.

\begin{figure}[H]
    \centering
    \includegraphics[width = 2.8in]{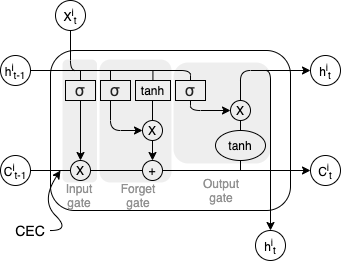}
    \caption{Block diagram of an LSTM cell.}
    \label{fig:lstm_cell}
\end{figure}

Figure \ref{fig:lstm_cell} illustrates the memory cell (neuron) of an LSTM whose principal characteristic is a constant error flow with few linear interactions through the constant error carousel (CEC) \cite{colah}. Using this carousel, the cell state vector or long-term memory, $C_t^i$, travels through the different timesteps with almost no change. The memory cell is a combination of gates, which, in the end, are MLP neurons created to make any change needed inside the memory cell, storing the information between timesteps nearly unchanged (Long-term memory) \cite{hochreiter}. The input gate selects which values will be updated for the next timestep; the forget gate is in charge of removing the irrelevant information; it deletes the values that will cause perturbations while maintaining the useful data; the output gate is responsible for sending the hidden state to the next timestep and the output state vector.

An LSTM has two activation functions. The named activation function has a similar behaviour to the activation function of MLPs. It is applied to the hidden states, i.e. to the information passed to the outside of the neuron. The recurrent activation function is applied to the input, forget and output gates of the memory cell.  
 
After defining the NN's structure, it is time for the training stage. We use supervised training, in which the NN is fed with a dataset subset, called a training dataset, and a learning algorithm changes the NN's parameters (weights and biases) to bring the outcome of the NN closer to the actual training dataset output. The loss function gives the parameter from which the algorithm measures the distance of the training predictions from the real values. Typically, the most common loss function is the mean squared error of the training prediction versus actual values, and we use it except when explicitly stated otherwise. 

Once the NN has learned, it has to be tested against new data to measure the NN's generalization capability for previously unseen data. To do so, the NN is now fed with a new data subset, the test dataset, not used in the training stage, and the NN's outcomes are compared with those in the test dataset.

\section{Models description} \label{sec:models}
In this section, we explain all the models studied and their adaptations to suit the conditions we placed on the study. The models have been selected attending to two criteria. On the one hand, that they return good predictions, and, on the other hand, that they are a variety representation of the most used NN models and architectures for BG prediction.

\begin{figure} [H]
    \centering
    \includegraphics[width=0.6\linewidth]{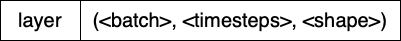}
    \caption{NN block diagram building block.}
    \label{fig:keras_plot}
\end{figure}

NN architecture is usually depicted using a standard graphical representation in which each NN's layer is described by a set of boxes, with one box for each layer and a set of arrows indicating the connections between the layers. The boxes representing each layer indicate first the type of layer and then two rows summarise the input and output parameter arrays. Each array can have two or three parameters. \Cref{fig:keras_plot} shows the case of having three parameters: the first parameter, \texttt{<batch>}, describes the batch size, i.e. the number of samples that will be introduced for each training loop, and the value `None' refers to a non-predefined dynamic batch size; the second parameter \texttt{<timesteps>} is the number of timesteps in RNNs, which is the number of previous samples in time that the layer processes; and the last one, \texttt{<shape>}, is the number of components in the input or output vector of the layer. In LSTM NNs,  \texttt{<shape>} refers to the number of units per layer, whereas in CNNs, \texttt{<shape>} is the number of filters per layer. In those layers with only two parameters, they are \texttt{<batch>} and \texttt{<shape>}.  

We also propose the use of NN ensemble models to predict BG. An ensemble model \cite{Dietterich2000} is a set of models trained either with different algorithms or datasets whose output is  aggregated to improve the quality of the predictions. The purpose of this technique is to reduce the generalization error of the prediction. To this aim, the base models in the ensemble have to be diverse and independent \cite{KOTU2015}. In this work, independence and diversity are attained by using different NN architectures. The choice of the models in the ensemble is made using the best models from the ten NNs evaluated in this work and they were selected according to the criteria in \cite{Calvo2018} explained in \Cref{sec:results}. The ensemble's outcome is the average of the individual model outcomes.

\subsection{Mirshekarian, 2017}

In \cite{mirs}, Mirshekarian \etal develop an LSTM NN for blood glucose prediction. \Cref{Mirsearch} summarizes the architecture parameters. It consists of just one hidden layer with five units, followed by a dense layer with one unit acting as the output layer. The recurrent activation function for the LSTM layer is the sigmoid function, and the activation function is $\tanh$. The activation function of the output neuron is the linear function. 

\begin{table} [H]
    \begin{center}
	\caption{Mirshekarian's architecture hyperparameters \cite{mirs}.}
	\label{Mirsearch}
	\begin{tabular}{c l c}
	    \toprule
	       Layer & Hyperarameter & Value \\
        \midrule
			\multirow{5}{*}{\rotatebox[origin=c]{90}{Hidden}} & Type & LSTM \\
			            & Layers & 1 \\
			            & Units per layer & 5 \\ 
                        & Recurrent activation function & sigmoid \\ 
						& Activation function & tanh  \\ 
		\midrule
    		\multirow{2}{*}{\rotatebox[origin=c]{90}{Out}} 
    		& Units  & 1 \\ 
			& Activation function & linear\\
		\midrule
		    & Number of parameters & 206 \\
		\bottomrule
	\end{tabular}
	\end{center}
\end{table}

\begin{figure} [H]
    \centering
    \includegraphics[width=0.35\linewidth]{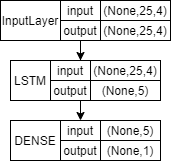}
    \caption{Mirshekarian's architecture block diagram.}
    \label{fig:mirsharch}
\end{figure}

\Cref{fig:mirsharch} illustrates the architecture. The input layer has four units, since, in the input layer, the shape is the number of features, with 25 timesteps, and its purpose is to receive the dataset to be inputted in the model. Next, there is the LSTM layer with five units with 25 timesteps, followed by the dense output  layer formed by a single dense neuron with five inputs and one output.

\subsection{Meijner, 2017}\label{sec:meijner}

Meijner's thesis \cite{meijner} presents two NNs with similar structures. Nevertheless, we pay attention to his LSTM-2 model because it is a modification of a standard LSTM whereas the other model is a standard LSTM. It operates under the assumption that BG can be predicted using a normal probability density function with mean $\mu$ and variance $\sigma^2$, which completely define the normal density function, and which are calculated by two parallel dense layers. Hence, for each input value, the $\mathbf{x}_t$, LSTM-2 model provides an estimation of the mean, $\mu_t$, and variance, $\sigma^2_t$. The loss function calculates the error of a miss-estimation of the probability density function parameters as the mean, over the $k$ values of the batch, of the negative logarithm of the actual glucose value's probability, $y_{t}$, when the NN estimation of the normal distribution parameters is $\mu_t$ and $\sigma_{t}^{2}$, that is, $\mathcal{N} \left ( y_{t}|\mu_{t},\sigma_{t}^{2} \right )$. Hence, when $y_t = \mu_t$ then  $\mathcal{N} \left ( y_{t}|\mu_{t},\sigma_{t}^{2} \right ) =1 $ and $\log \left ( \mathcal{N} \left ( y_{t}|\mu_{t},\sigma_{t}^{2} \right ) \right ) =0$ whereas if $y_t$ is far from $\mu_t$ then $\log \left ( \mathcal{N} \left ( y_{t}|\mu_{t},\sigma_{t}^{2} \right ) \right )$ has a high value, increasing the loss function.

\Cref{table:meijarch} summarizes the architecture of this NN. It consists of one LSTM layer with four units, each one corresponding to a different feature, followed by two parallel dense layers composed of one unit each; one returns the predicted value (mean), and the other outputs a confidence interval (standard deviation).

\begin{table} [H]
     \begin{center}
 	\caption{Meijner's architecture hyperparameters \cite{meijner}.}
 	\label{table:meijarch}
 	\begin{tabular}{c l c}
 	    \toprule
             & Loss function &  
             \(\displaystyle \frac{-\sum_{t=1}^{k} \log \left ( \mathcal{N} \left ( y_{t}|\mu_{t},\sigma_{t}^{2} \right ) \right )}{k} \) \label{dd}\\
         \midrule
         	 Layer & Hyperparameter & Value \\
         \midrule
 			\multirow{5}{*}{\rotatebox[origin=c]{90}{Hidden}} & Type & LSTM \\
 			            & Layers &  1\\
 			            & Units per layer & 4 \\ 
                        & Recurrent activat. function & sigmoid\\ 
 						& Activation function & tanh  \\ 
 		\midrule
     		\multirow{3}{*}{\rotatebox[origin=c]{90}{Output}} & Units  & 2 (mean  and  variance) \\ 
 			& Activat. function mean & linear \\
 			& Activat. function variance & ELU, see \Cref{eq:ELU}\\
 		\midrule
		    & Number of parameters & 154 \\
 		\bottomrule
 	\end{tabular}
 	\end{center}
\end{table}

\begin{figure} [H]
\centering
        \includegraphics[width=0.55\linewidth]{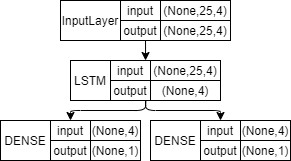}
        \caption{Meijner's architecture block diagram.}
        \label{fig:meijnerarch}
\end{figure}

\Cref{fig:meijnerarch} illustrates the architecture. The input layer has four units with 25 timesteps. Next, the LSTM layer has four units and 25 timesteps followed by two dense output layers, for the mean and the variance, formed by dense units with four inputs and one output.

\subsection{G\"ulesir, 2018}

In \cite{gizem}, G\"ulesir \etal propose a CNN to forecast BG. The two aforementioned models use LSTM, the state-of-the-art NN for timeseries forecasting, so the main contribution and most important difference in this paper is the application of CNNs to the BG forecasting problem. The authors intend the timesteps in the timeseries to be a one-dimensional image. The set of values of the incoming input in a sample will correspond to a pixel of such an image, and the three-colour combination in a pixel (a combination of red, green and blue) is now every incoming feature in the timeseries, i.e., BG, CH, basal insulin, and insulin bolus.

\begin{figure} [H]
    \centering
    \includegraphics[width=0.40\linewidth]{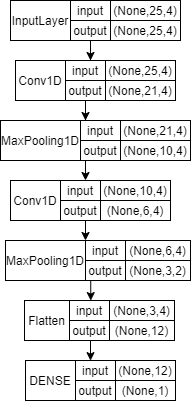}
    \caption{G\"ulesir's architecture block diagram.}
    \label{gulesirarch}
\end{figure}

\Cref{gulesirarch} illustrates their proposal. This CNN has two convolutional layers. As in a typical CNN, each one is followed by a max-pooling layer whose objective is to summarize the patterns detected to have more flexibility when detecting similar patterns for future timesteps. As a consequence of this summary, the output's dimensionality is reduced. Finally, a flatten layer is introduced to create a one-dimensional array because the dense output layer can only read one-dimensional vectors.

\Cref{tab:gulesir} summarizes the key parameters of the NN layers. Both convolutional layers are designed in the same way. The number of filters is four and the size of each filter is set to five. The max-pooling layer that follows each CNN layer has a pool size with a value of two, which causes its output vector to be halved with respect to the input one.

 \begin{table} [H]
     \begin{center}
 	\caption{Gülesir's architecture hyperparameters \cite{gizem}.}
 	\label{tab:gulesir}
 	\begin{tabular}{c l c}
 	    \toprule
         	 Layer & Hyperarameter & Value \\
         \midrule
 			\multirow{8}{*}{\rotatebox[origin=c]{90}{Hidden}} & Type & Conv \\
 			            & Layers &  2\\
                        & Number of Filters & 4, 4\\
                        & Filter size & 5, 5\\
 						& Activation function & ReLU \\ 
 						\cmidrule{2-3}
 						& Type & MaxPooling1D \\
 			            & Layers &  2\\
 			            & Pool size & 2, 2 \\ 
 		\midrule
     		\multirow{2}{*}{\rotatebox[origin=c]{90}{Out}} 
     		        & Units  & 1 \\ 
			        & Activation function & linear\\
		\midrule
		    & Number of parameters & 181 \\
		 \bottomrule
 	\end{tabular}
 	\end{center}
 \end{table}

\subsection{Sun, 2018}

\begin{figure}[H]
    \centering
    \includegraphics[width=3.2in]{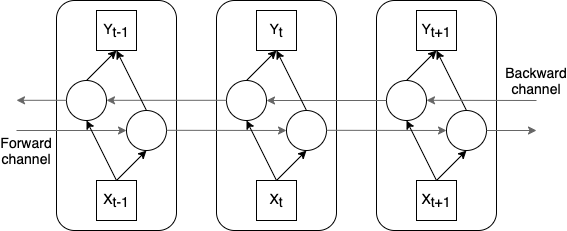}
    \caption{Bidirectional LSTM NN.}
    \label{brnn}
\end{figure}

In \cite{sun}, Sun \etal present a Bi-LSTM to predict BG values. The availability of a full timeseries enables access to future data with respect to a given timestep. This availability is used by Bi-LSTM NNs. \Cref{brnn} illustrates the block diagram of such an NN. It has two input channels; in the forward channel, the timewindow of the past \SI{120}{\min} is processed forward in time, i.e., starting from the furthest timestep to the current timestep, whereas in the backwards channel this timewindow is processed from the current timestep to the furthest one.

\begin{figure} [H]
    \centering
    \includegraphics[width=0.40\linewidth]{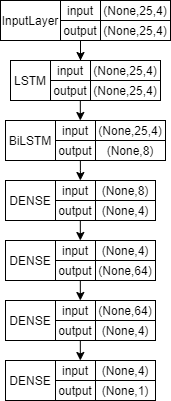}
    \caption{Sun's Bi-LSTM NN architecture block diagram.}
    \label{fig:sunarch}
\end{figure}

In the article, the authors do not introduce the activation functions and some parameters of the model, but only the number of units and their type. For this reason, we devised the Bi-LSTM NN depicted in \Cref{fig:sunarch} using the state-of-the-art parameters for every layer in the model, which leads to some differences, i.e., our output vector in the bidirectional layer has a shape of eight units while in Sun's work it is set to four units.

\Cref{tab:sun} presents the NN layers' hyperparameters. Merging is the mode by which the forward and backward channels of the Bi-LSTM are combined, which in this case are concatenated to generate the output vector in the layer.

\begin{table} [H]
    \begin{center}
	\caption{Sun's architecture hyperparameters \cite{sun}.}
	\label{tab:sun}
	\begin{tabular}{c l c}
	    \toprule
	       Layer & Hyperparameter & Value \\
        \midrule
			\multirow{15}{*}{\rotatebox[origin=c]{90}{Hidden}} & Type & LSTM \\
			            & Layers & 1 \\
			            & Units per layer & 4 \\ 
                        & Recurrent activation function & sigmoid \\ 
						& Activation function & tanh  \\ 
						\cmidrule{2-3}
						& Type & BiLSTM \\
			            & Layers & 1 \\
			            & Units per layer & 4 \\ 
                        & Recurrent activation function & sigmoid \\ 
						& Activation function & tanh  \\ 
						& Merging & Concatenation \\
						\cmidrule{2-3}
						& Type & Dense \\
			            & Layers & 3 \\
			            & Units per layer & 4, 64, 4 \\ 
                        & Activation function & linear\\ 
		\midrule
    		\multirow{2}{*}{\rotatebox[origin=c]{90}{Out}} 
    		& Units  & 1 \\ 
			& Activation function & linear\\ 
		\midrule
		    & Number of parameters & 1053 \\
		\bottomrule
	\end{tabular}
	\end{center}
\end{table}

\subsection{Idriss, 2019}

\begin{figure} [H]
    \centering
    \includegraphics[width=0.40\linewidth]{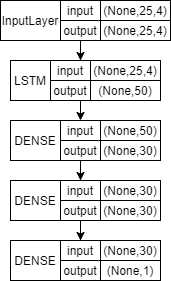}
    \caption{Idriss' architecture block diagram.} \label{fig:idrissarch}
\end{figure}

In \cite{idriss}, Idriss \etal propose the model in \Cref{fig:idrissarch}; a model with one LSTM layer to study the temporal dimension of the data and two dense layers to extract the remaining features of BG dynamics.  

\Cref{tab:idriss} presents the architecture hyperparameters on a per-layer basis. We selected the number of units in each layer according to the best option proposed in the article. Idriss tested different unit combinations, obtaining better results with 50 units in each LSTM layer and 30 units in each dense layer. As there is no information on each unit's activation functions, we set the most common activation functions in each layer: the sigmoid function for the dense layer and the recurrent activation function of the LSTM, and the $\tanh$ function for the LSTM's activation function.

\begin{table} [H]
    \begin{center}
	\caption{Idriss's architecture hyperparameters \cite{idriss}.}
	\label{tab:idriss}
	\begin{tabular}{c l c}
	    \toprule
	       Layer & Hyperparameter & Value \\
        \midrule
			\multirow{9}{*}{\rotatebox[origin=c]{90}{Hidden}} 
			            & Type & LSTM \\
			            & Layers & 1 \\
			            & Units per layer & 50 \\ 
                        & Recurrent activation function & sigmoid \\ 
						& Activation function & tanh  \\ 
						\cmidrule{2-3}
						& Type & Dense \\
			            & Layers & 2 \\
			            & Units per layer & 30, 30 \\ 
                        & Activation function & linear\\ 
		\midrule
    		\multirow{2}{*}{\rotatebox[origin=c]{90}{Out}} 
    		& Units  & 1 \\ 
			& Activation function & linear\\ 
		\midrule
		    & Number of parameters & 13491 \\
		\bottomrule
	\end{tabular}
	\end{center}
\end{table}

\subsection{Aiello, 2019} 

In \cite{aiello}, Aiello \etal propose an LSTM model that uses a timewindow of the last \SI{120}{\min} of data and, additionally, a timewindow of \SI{30}{\min} of data of the features with known future values, such as basal insulin or insulin bolus, or estimated values such as CH intakes.

\begin{figure} [H]
    \centering
    \includegraphics[width=0.75\linewidth]{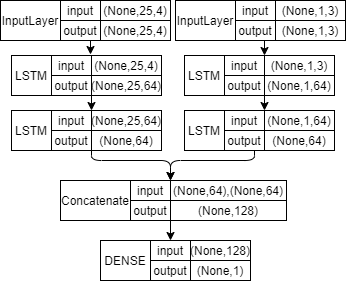}
    \caption{Aiello's architecture block diagram.}
    \label{fig:aielloarc}
\end{figure}

\Cref{fig:aielloarc} shows the two-path model in which future and test data is processed separately by two LSTM layers, with 64 units each; the branch of past data has 25 timesteps while the future data branch has one timestep of \SI{30}{\min}. As is normally the case, the recurrent activation function is the sigmoid, and the activation function is tanh. Then, both branches are concatenated before returning the final output.

\Cref{tab:aiello} presents the architecture's hyperparameters on a per-layer basis according to the values presented in their paper.

\begin{table} [H]
    \begin{center}
	\caption{Aiello's architecture hyperparameters \cite{aiello}.}
	\label{tab:aiello}
	\begin{tabular}{c l c}
	    \toprule
	       Layer & Hyperparameter & Value \\
        \midrule
			\multirow{5}{*}{\rotatebox[origin=c]{90}{Hidden}} 
			            & Type & LSTM \\
			            & Layers & 4 \\
			            & Units per layer & 64, 64, 64, 64 \\ 
                        & Recurrent activation function & sigmoid \\ 
						& Activation function & tanh  \\ 
		\midrule
    		\multirow{2}{*}{\rotatebox[origin=c]{90}{Out}} 
    		& Units  & 1 \\ 
			& Activation function & linear\\ 
		\midrule
		    & Number of parameters & 101249 \\
		\bottomrule
	\end{tabular}
	\end{center}
\end{table}

\subsection{Zhu, 2020}

\begin{figure}[H]
    \centering
    \includegraphics[width=3.4in]{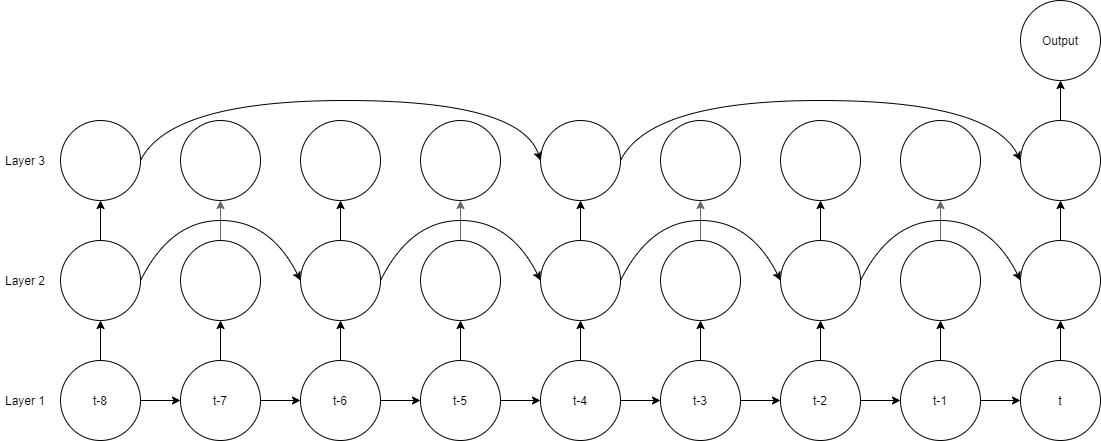}
    \caption{Zhu Dilated RNN architecture.}
    \label{fig:dilatedrnn}
\end{figure}

\begin{figure} [H]
    \centering
    \includegraphics[width=0.40\linewidth]{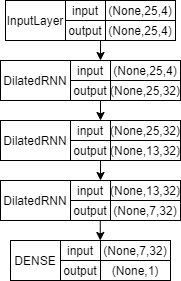}
    \caption{Zhu's architecture block diagram.} \label{fig:zhuarch}
\end{figure}

In \cite{zhu}, Zhu \etal present a dilated RNN. Dilation consists of skipping some steps according to the dilation rate to reduce the number of parameters and obtain greater efficiency while eliminating redundant information. \Cref{fig:dilatedrnn} illustrates this technique, which is commonly used for CNNs, and how the authors have applied the dilation of the layers in RNNs. \Cref{fig:zhuarch} shows the NN architecture in its standard format. It has three RNN dilation  layers; the first layer has a dilation rate of 1 and works as a conventional RNN. Then, the dilation rate is multiplied by 2 from one layer to the next; they have a value of 2 and 4, respectively. 

\Cref{tab:zhu} presents the NN's hyperparameters. Following their article, we use vanilla RNN units since these have demonstrated the best performance. The number of units in each layer is 32 according to the configuration with the best results, and the activation function is  tanh. Note that, as they are RNN and not LSTM units, they do not have a recurrent activation function. Finally, a dense layer with one unit outputs the predicted BG value.

\begin{table} [H]
    \begin{center}
	\caption{Zhu's architecture hyperparameters \cite{zhu}.}
	\label{tab:zhu}
	\begin{tabular}{c l c}
	    \toprule
	       Layer & Hyperparameter & Value \\
        \midrule
			\multirow{5}{*}{\rotatebox[origin=c]{90}{Hidden}} 
			            & Type & Dilated RNN \\
			            & Layers & 3 \\
			            & Units per layer & 32, 32, 32 \\ 
						& Activation function & tanh  \\ 
						& Dilation rate & 1, 2, 4 \\
		\midrule
    		\multirow{2}{*}{\rotatebox[origin=c]{90}{Out}} 
    		& Units  & 1 \\ 
			& Activation function & linear\\ 
		\midrule
		    & Number of parameters & 5377 \\
		\bottomrule
	\end{tabular}
	\end{center}
\end{table}

\subsection{Mayo, 2020}

In \cite{mayo}, Mayo and Koutny address this problem using a different approach. Instead of treating BG prediction as a timeseries forecast problem, they consider it as a classification problem. The fluctuations of the predicted BG levels have a different impact on the patient depending on the glucose levels' actual value; it is not the same to have an error of \SI{10}{\BGunit} when the patient is in a hypoglycemia event (BG \textless\  \SI{60}{\BGunit}) as when the patient has euglycemia (BG between \SIrange[range-phrase=\text{ and }]{70}{160}{\BGunit}). In the second scenario, the patient does not suffer from any repercussion in their health, while in the first case, the patient can suffer a severe health threat if not treated. To deal with this phenomenon, the authors preprocess BG levels using the risk domain transform, \Cref{risk}, a nonlinear function whose output, $r$, spans the range $[-2,2]$ and whose normoglycemic measurements lie in the range $[-0.9, 0.9]$ \cite{Kovatchev2000}. Using $r$, the hypo-\- and hyperglycemic ranges have equal size and significance, minimizing the chance of bias in statistical analysis, e.g., due to larger absolute error sizes in the hyperglycemic range. Next, the authors divided the risk range into 100 equally spaced bins to define a set of classes with sufficient precision for the predictions.

\begin{equation}
    \label{risk}
    r(\textrm{bg}_{t}) =  1.509(\log(\textrm{bg}_{t})^{1.084}-5.381)
\end{equation}

\begin{figure} [H]
    \centering
    \includegraphics[width=0.40\linewidth]{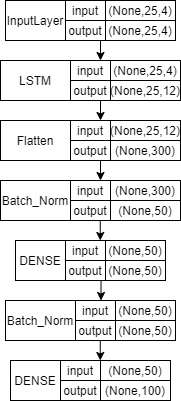}
    \caption{Mayo's architecture block diagram.}     \label{fig:mayoNN}
\end{figure}

Once the blood glucose is preprocessed, it is time for the NN model. \Cref{fig:mayoNN} illustrates the NN architecture and \Cref{tab:mayo} presents the layers' hyperparameters. It consists of an LSTM with 12 units, the activation function is  tanh and the recurrent activation function is the sigmoid. It is followed by a flatten layer and a batch normalization layer to avoid overfitting. Then, a dense layer processes the information with 50 units using the ReLU activation function. This layer is followed by another batch normalization layer. Finally, the output layer has 100 units to address the different classes previously defined with a linear activation function.

\begin{table} [H]
    \begin{center}
	\caption{Mayo's architecture hyperparameters \cite{mayo}.}
	\label{tab:mayo}
	\begin{tabular}{c l c}
	    \toprule
	       Layer & Hyperparameter & Value \\
        \midrule
			\multirow{9}{*}{\rotatebox[origin=c]{90}{Hidden}} 
			            & Type & LSTM \\
			            & Layers & 1 \\
			            & Units per layer & 12 \\ 
			            & Recurrent activation function & sigmoid \\
						& Activation function & tanh  \\ 
						\cmidrule{2-3}
						& Type & Dense \\
			            & Layers & 1 \\
			            & Units per layer & 50 \\ 
						& Activation function & ReLU  \\ 
		\midrule
    		\multirow{2}{*}{\rotatebox[origin=c]{90}{Out}} 
    		& Units  & 100 \\ 
			& Activation function & linear\\
		\midrule
		    & Number of parameters & 22366 \\
		\bottomrule
	\end{tabular}
	\end{center}
\end{table}

\subsection{Muñoz, 2020}

\begin{figure*}
    \centering
    \includegraphics[width=1\linewidth]{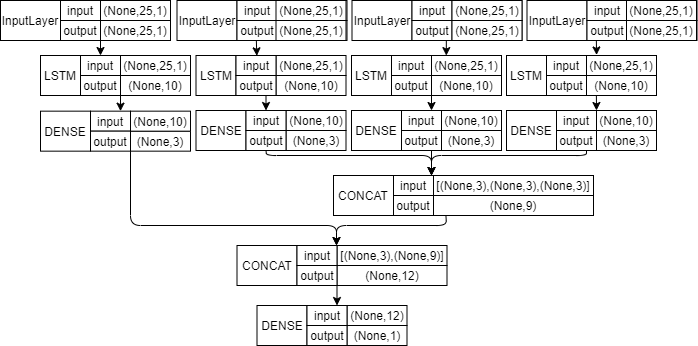}
    \caption{Muñoz's architecture block diagram.} \label{fig:munozNN}
\end{figure*}

In \cite{munoz}, Muñoz designed an NN to mimic the metabolic behavior of physiological BG models. His idea was to create a neural network capable of learning the process that models the digestion of CH and the absorption of insulin,  combined with the data history of BG levels. \Cref{fig:munozNN} represents the NN architecture, and \Cref{tab:munoz} presents the architecture's hyperparameters. The system has four submodels, as many as features to be processed. Each submodel consists of an LSTM layer with ten units,  the recurrent sigmoid activation function and the ReLU activation function, followed by a dense layer with three units, and a ReLU activation function. 

\begin{table} [H]
    \begin{center}
	\caption{Muñoz's architecture hyperparameters \cite{munoz}.}
	\label{tab:munoz}
	\begin{tabular}{c l c}
	    \toprule
	       Layer & Hyperparameter & Value \\
        \midrule
			\multirow{9}{*}{\rotatebox[origin=c]{90}{Hidden}} 
			            & Type & LSTM \\
			            & Layers & 4 \\
			            & Units per layer & 10, 10, 10, 10 \\ 
			            & Recurrent activation function & sigmoid \\
						& Activation function & ReLU  \\ 
						\cmidrule{2-3}
						& Type & Dense \\
			            & Layers & 4 \\
			            & Units per layer & 3, 3, 3, 3 \\ 
						& Activation function & ReLU  \\ 
		\midrule
    		\multirow{2}{*}{\rotatebox[origin=c]{90}{Out}} 
    		& Units  & 1 \\ 
			& Activation function & linear \\
		\midrule
		    & Number of parameters & 2075 \\
		\bottomrule
	\end{tabular}
	\end{center}
\end{table}

After each feature is processed separately, the CH and insulin rates are concatenated together, returning a prediction without BG dynamics to test how the model works without past BG information. Then, this information is concatenated with BG levels to predict the final values. In this paper, we test the model using all the features because we want to compare models under the same conditions.

\subsection{Khadem, 2020}

In \cite{khadem}, Khadem \etal propose a system which is a combination of six models, called base-learners. The six models are two LSTM models, two dense models, and two Partial Least Square Regression (PLSR) models where PLSR is a very popular basic linear regression optimized for predictions \cite{plsr}  because of the ease of implementation and its minimal computational time.

\begin{figure*}
    \centering
    \includegraphics[width=1\linewidth]{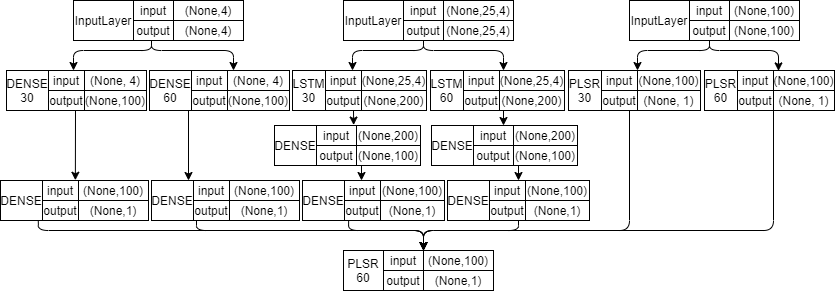}
    \caption{Khadem's architecture block diagram.} \label{fig:khademNN}
\end{figure*}

As \Cref{fig:khademNN} illustrates, three of them were trained with \SI{30}{\min} horizon predictions and the remainder with \SI{60}{\min} horizon predictions. 

\Cref{tab:khadem} details the layers' hyperparameters. The Dense model is a 1-layer dense NN with 100 units and with ReLU as activation function. The LSTM model consists of an LSTM layer with 200 units, 25 timesteps, four input features, recurrent sigmoid activation function, and ReLU activation function, and a dense layer with 100 units and ReLU activation function. Finally, a PLSR receives all the base-learners' output, acting as meta-learner, to decide the final prediction. 

\begin{table} [H]
    \begin{center}
	\caption{Khadem's architecture hyperparameters \cite{khadem}.}
	\label{tab:khadem}
	\begin{tabular}{c l c}
	    \toprule
	       Layer & Hyperparameter & Value \\
        \midrule
			\multirow{9}{*}{\rotatebox[origin=c]{90}{Hidden}} 
			            & Type & LSTM \\
			            & Layers & 2 \\
			            & Units per layer & 200, 200 \\ 
			            & Recurrent activation function & sigmoid \\
						& Activation function & ReLU  \\ 
						\cmidrule{2-3}
						& Type & Dense \\
			            & Layers & 4 \\
			            & Units per layer & 100, 100, 100, 100 \\ 
						& Activation function & ReLU  \\ 
						\cmidrule{2-3}
					    & Type & PLSR \\
					    & Layers & 2 \\
		\midrule
    		\multirow{6}{*}{\rotatebox[origin=c]{90}{Output}} 
						& Type & Dense \\
			            & Layers & 4 \\
			            & Units per layer & 1, 1, 1, 1 \\ 
						& Activation function & ReLU  \\ 
						\cmidrule{2-3} 
						& Type & PLSR \\
					    & Layers & 1 \\
		\midrule
		    & Number of parameters & 369810 \\
		\bottomrule
	\end{tabular}
	\end{center}
\end{table}

\section{Experiment Results} \label{sec:experimental_results}

For this paper, we replicate the ten NNs described in \Cref{sec:models} using Python 3.7, Tensorflow 2.2.0 and Keras 2.3.1. These models use the OhioT1DM dataset \cite{ohio} for training and testing each model. The OhioT1DM dataset has recently been used for the "Blood Glucose Prediction Challenge" of the "Workshop on Knowledge Discovery in Healthcare Data", which brings together around 20 models, both neural network and non-neural network models, and it is also used in the literature. In addition, this dataset has been used in at least five of the models we replicate in this paper. It can therefore be considered as one of the references for \textit{in vivo} data used for this research area. Thus, it is a suitable dataset to be used in a comparison. In particular, we use the second cohort of the OhioT1DM dataset, which contains six patients with five males and one female aged between 20 and 80 years who participated in an IRB-approved study for eight weeks each. They used Medtronic Enlite CGM sensors, reported life-event data via an app and provided physiological data using the Empatica Embrace fitness band.

\subsection{Data preprocessing}

To begin with, the data is preprocessed. BG timeseries sample time is \SI{5}{\min} and we use cubic splines in order to complete missing samples and create a timeseries compatible with the models, with a total amount of \num{92791} samples. We chose 5-minute timesteps for both  data acquisition and  backpropagation, with a history of \SI{120}{\min}, or 24 timesteps, for backpropagation plus the actual timestep, within each element of the dataset.

For this dataset, the input features are BG levels ($\mathrm{bg}$), basal insulin ($\mathrm{bas}$), insulin boluses ($\mathrm{bol}$), and CH intakes ($\mathrm{ch}$), so that $\mathbf{x}(t) = \left ( \mathrm{bg}(t), \mathrm{bas}(t), \mathrm{bol}(t), \mathrm{ch}(t) \right )$ is the input vector at time $t$. We chose these features because they have the highest impact on BG dynamics. In this work, BG levels are multiplied by a factor of \num{0.01} so the NN can reach BG prediction faster according to the algorithm learning rate and to submit them in a similar scale to the remaining three features. These features have been normalized within the range $[0.1]$ to increase the distance between the different values of the features to make it easier for the neural networks to appreciate a change within the feature for pattern identification.

\subsection{Training and testing}

We defined the same conditions to train all the NNs. The models with a pretraining phase are tested twice: with its pretraining and with the same training conditions as the rest of the models. We do not include the results with pretraining because the differences in the results are not relevant. The training and test data are split as provided by the OhioT1DM database, and training is performed using the 80/20 10-fold cross validation approach. \Cref{tab:samples} presents the number of samples each patient has for training and testing procedures.

\begin{table}[htp]
    \centering
        \caption{Number of training and testing samples per patient.}
    \begin{tabular}{ | l | c | c |}
    \toprule
        \multicolumn{1}{c}{} & \multicolumn{2}{c}{Number of samples} \\ \midrule
        Patient & Training & Testing \\ \midrule
        540 & 13062 & 3018 \\
        544 & 12624 & 3089 \\
        552 & 11050 & 3903 \\
        567 & 13489 & 2824 \\
        584 & 13201 & 2948 \\
        596 & 13582 & 2956 \\ \midrule
        \multicolumn{3}{c}{} \\
	\bottomrule
    \end{tabular}
    \label{tab:samples}
\end{table}

We bound the NN's predictions between \SIrange[range-phrase=\text{ and }]{40}{400}{\milli\gram \per \deci \liter} because the values in the dataset are already bounded since the readings come from CGMs whose minimum and maximum values are \SI{40}{\BGunit} and \SI{400}{\BGunit}, respectively.

The Adam algorithm \cite{adam} with a learning rate of \num{0.01} and the mean squared error, \Cref{eq:mse}, as loss function is used in the training. The training consists of 100 epochs with an early stopping of 10 epochs' patience. In the model validation, the mean absolute error, \Cref{eq:mae}, is applied as the metric function.
\subsection{Results} \label{sec:results}

We forecast the blood glucose at three prediction horizons, $\mathrm{PH}=\{30, 60, 120\}$ min; we take $\mathrm{ph}=$ \SI{30}{\min} as the starting short-term prediction horizon and double it progressively to define the medium- and long-term prediction horizons. We denote the actual BG value at time $t$ as $\mathrm{bg}(t)$, the actual future BG value $\mathrm{ph} \in \mathrm{PH}$ minutes ahead of time $t$ as $\BGact{ph} = \mathrm{bg}(t+ph)$, and the predicted BG $ph$ minutes ahead of time $t$ as $\widehat{\mathrm{bg}}_{ph}(t)$. The predictions are evaluated on a per-patient basis using the most common error metrics, \Crefrange{eq:mse}{eq:mard} respectively: mean squared error (MSE), root mean squared error (RMSE), mean absolute error (MAE), R-squared (R$^{2}$), correlation coefficient (CC), fit (Fit), and mean absolute relative difference (MARD). In them, $n$ is the number of predictions per patient, and $\overline{\mathrm{bg}}_{ph} = \frac{1}{n} \sum_{t=1}^n \BGact{ph}$ and $\overline{\widehat{\mathrm{bg}}}_{ph} = \frac{1}{n} \sum_{t=1}^n \BGpre{ph}$ are the mean values. 

\begin{align}
    \mathrm{MSE}_{ph}   &= \frac{1}{n} \sum_{t=1}^{n} \left (\BGpre{ph} - \BGact{ph} \right )^{2}                       \label{eq:mse} \\
    \mathrm{RMSE}_{ph}  &= \sqrt{\mathrm{MSE}_{ph}}                                                                        \label{eq:rmse} \\
    \mathrm{MAE}_{ph}   &= \frac{1}{n} \sum_{t=1}^{n} \lvert \BGpre{ph} - \BGact{ph} \rvert               \label{eq:mae} \\
    \mathrm{R}_{ph}^{2} &= 1 - \frac{\sum\limits_{t=1}^{n} \left ( \BGpre{ph} - \BGact{ph} \right )^{2}}{\sum\limits_{t=1}^{n} \left ( \BGact{ph} - \BGactAvg{ph} \right ) ^{2}}  \label{r2} \\
    \mathrm{CC}_{ph} &= \frac{\sum\limits_{t=1}^{n} \left ( \BGpre{ph} - \BGpreAvg{ph} \right ) \left ( \BGact{ph} - \BGactAvg{ph} \right )}{\sqrt{\sum\limits_{t=1}^{n} \left ( \BGpre{ph} - \BGpreAvg{ph} \right )^{2}\sum\limits_{t=1}^{n} \left ( \BGact{ph} - \BGactAvg{ph} \right )^{2}}}        \label{cc} \\
    \mathrm{Fit}_{ph} &= 1-\frac{\frac{1}{n}\sum\limits_{t=1}^{n}\left|\BGpre{ph} - \BGactAvg{ph} \right |}{\frac{1}{n}\sum\limits_{t=1}^{n} \left|\BGact{ph} - \BGactAvg{ph} \right |}  \label{eq:fit} \\
    \mathrm{MARD}_{ph} &= \frac{1}{n}\cdot\sum_{t=1}^{n}\frac{|\BGpre{ph} - \BGact{ph}|}{\BGact{ph}}        \label{eq:mard}
\end{align}

Tables \ref{tab:performance30}, \ref{tab:performance60} and \ref{tab:performance120} show the results of the 10-fold cross validation over the different prediction horizons. The results are the average of the metrics over the patients with the standard error of the mean deviation. Green cells in each column highlight the model with the best performance, whereas the grey coloured cells are the worst. We have to differentiate two points of view to analyse these metrics: first, the point of view of the predictive artificial intelligence  tool from which we only take into account the results and do not look at the behaviour of BG levels; secondly, the clinical point of view in which we observe how the predictions affect the patients.

As an example, we present the RMSE values, as we can draw the same conclusions if we analyse the MSE or the MAE. For $\mathrm{ph}=\SI{30}{\min}$, there is a difference of only \SI{3.84}{\BGunit}, between the models with lowest (Sun, 2018) and highest (Idriss, 2019) RMSE$_{30}$. For $\mathrm{ph}=$\SI{60}{\min}, the difference is slightly higher, \SI{9.99}{\BGunit}, between the Sun model and the Aiello one. Finally, the lowest RMSE$_{120}$ is \SI{54.70}{\BGunit} for the Muñoz model and the highest one is \SI{65.79}{\BGunit} for the Idriss one. Hence, these differences can be notable in a numerical analysis of the results but are irrelevant from the clinical point of view.

R$^{2}$ can be interpreted as the explainability of the model, i.e., how much of the data can be explained by each of the models. For $\mathrm{ph}=\SI{30}{\min}$, even the worst model explains \SI{85}{\percent} of the data variability; thus, the predictions of all the models are promising. For $\mathrm{ph}=$\SI{60}{\min}, the explainability of the models remains high enough, explaining between \SIrange[range-phrase ={ and }]{45}{67}{\percent} of the data variability. But, for $\mathrm{ph}=$\SI{120}{\min}, only \SI{16}{\percent} of the data variability can be explained. 

Regarding CC, we compare predictions versus actual values, so the highest performance corresponds to values near 1. CC$_{30}$ values are around \num{0.94} and CC$_{60}$ values range from \num{0.72} to \num{0.83} and are still relevant figures, while the highest CC$_{120}$ is \num{0.48}, indicating a poor fit for the prediction. In addition, $\mathrm{Fit}_{30}$ and $\mathrm{Fit}_{60}$ values show once again that all the models predict with very similar values. Finally, for $\mathrm{Fit}_{120}$, we find values near zero, or even negative; this means that the predictions are further from the mean than the targets or, in other words, they do not predict correctly. 

Finally, MARD is the most common metric used to analyse the accuracy of CGM systems \cite{danne}. It measures the difference between the actual values and the predicted ones. So, the lower the MARD is the more accurate the predictions are. For $\mathrm{ph}=\SI{30}{\min}$ we find a difference between \num{0.10} and \num{0.12} which indicates a good correlation of the predictions, for $\mathrm{ph}=\SI{60}{\min}$ the difference is greater, as expected, between \num{0.18} and \num{0.26}.  and, for $\mathrm{ph}=\SI{120}{\min}$ MARD values lie between \num{0.32} and \num{0.38}.

\begin{table*}[htp!]
    \centering
        \caption{Results of the ten NN models for each performance metric for $\mathrm{ph}=$\SI{30}{\min}. Green cells in each column highlight the model with the best performance, whereas the grey coloured cells are the worst.}

    \resizebox{14cm}{!}{\begin{tabular}{ | l | c | c | c | c | c | c | c | }
    \toprule
    \multicolumn{1}{c}{}   & \multicolumn{7}{c}{Metrics}             \\  \cmidrule{2-8}  
	\multicolumn{1}{c|}{} & RMSE $\left ( \si{\BGunit} \right )$            & MSE  $\left ( \si{\square\milli\gram\per\square\deci\liter} \right )$                & MAE  $\left ( \si{\BGunit} \right )$            & R$^2$         & CC            & FIT           & MARD          \\ \midrule
	Mirshekarian          & \num{21.34(171)} & \num{470.08(7248)}  & \num{15.38(128)} & \num{0.87(2)} & \num{0.94(1)} & \num{0.68(3)} & \cellcolor{green}\num{0.10(1)} \\ 
	Meijner               & \num{19.92(135)} & \num{405.79(5326)}  & \cellcolor{green}\num{14.21(95)}  & \cellcolor{green}\num{0.89(3)} & \cellcolor{green}\num{0.95(2)} & \cellcolor{green}\num{0.71(4)} & \num{0.10(1)} \\ 
	Gülesir               & \num{22.18(132)} & \num{500.66(5867)}  & \num{16.52(101)} & \num{0.86(1)} & \cellcolor{gray}\num{0.93(1)} & \cellcolor{gray}\num{0.66(1)} & \num{0.11(1)} \\ 
	Sun                   & \cellcolor{green}\num{19.73(131)} & \cellcolor{green}\num{397.76(5153)}  & \num{14.54(92)}  & \cellcolor{green}\num{0.89(1)} & \cellcolor{green}\num{0.95(1)} & \num{0.70(1)} & \cellcolor{green}\num{0.10(1)} \\ 
	Idriss                & \cellcolor{gray}\num{23.57(198)} & \cellcolor{gray}\num{574.97(9183)}  & \cellcolor{gray}\num{16.59(126)} & \cellcolor{gray}\num{0.85(2)} & \cellcolor{gray}\num{0.93(1)} & \cellcolor{gray}\num{0.66(2)} & \cellcolor{gray}\num{0.12(1)} \\ 
	Aiello                & \num{22.64(176)} & \num{527.84(8021)}  & \num{15.89(119)} & \num{0.86(2)} & \cellcolor{gray}\num{0.93(1)} & \num{0.67(2)} & \num{0.11(1)} \\ 
	Zhu                   & \num{21.74(145)} & \num{482.95(6332)}  & \num{15.93(110)} & \num{0.87(1)} & \num{0.94(1)} & \num{0.67(1)} & \num{0.11(1)} \\ 
	Mayo                  & \num{22.35(248)} & \num{530.22(11534)} & \num{14.99(152)} & \num{0.86(2)} & \cellcolor{gray}\num{0.93(1)} & \num{0.69(3)} & \cellcolor{green}\num{0.10(1)} \\ 
	Muñoz                 & \num{21.22(139)} & \num{460.00(5945)}  & \num{15.74(98)}  & \num{0.88(1)} & \num{0.94(1)} & \num{0.67(2)} & \cellcolor{gray}\num{0.12(1)} \\ 
	Khadem                & \num{21.80(156)} & \num{487.51(6531)}  & \num{15.23(115)} & \num{0.86(2)} & \num{0.94(1)} & \num{0.68(3)} & \num{0.11(1)} \\ \midrule
	\multicolumn{8}{c}{} \\
	\bottomrule
    \end{tabular}}
    \label{tab:performance30}
\end{table*}

\begin{table*}[htp]
    \centering
    \setlength\tabcolsep{2pt}
        \caption{Results of the ten NN models for each performance metric for $\mathrm{ph}=$\SI{60}{\min}. Green cells in each column highlight the model with the best performance, whereas the grey coloured cells are the worst.}

\resizebox{14cm}{!}{\begin{tabular}{ | l | c | c | c | c | c | c | c | }
    \toprule
    \multicolumn{1}{c}{}   & \multicolumn{7}{c}{Metrics}             \\  \cmidrule{2-8}  
	\multicolumn{1}{c|}{} & RMSE  $\left ( \si{\BGunit} \right )$           & MSE $\left ( \si{\square\milli\gram\per\square\deci\liter} \right )$                  & MAE  $\left ( \si{\BGunit} \right )$            & R$^2$         & CC            & FIT           & MARD          \\ \midrule 
	Mirshekarian          & \num{38.58(280)} & \num{1527.85(21132)} & \num{28.58(203)} & \num{0.58(5)} & \num{0.79(3)} & \num{0.41(4)} & \num{0.20(1)} \\ 
	Meijner               & \num{36.55(254)} & \num{1368.13(18837)} & \num{26.67(178)} & \num{0.63(3)} & \num{0.81(2)} & \cellcolor{green}\num{0.45(3)} & \num{0.19(1)} \\ 
	Gülesir               & \num{37.25(242)} & \num{1387.56(18084)} & \num{28.46(183)} & \num{0.62(3)} & \num{0.80(2)} & \num{0.41(2)} & \num{0.20(1)} \\ 
	Sun                   & \cellcolor{green}\num{34.48(212)} & \cellcolor{green}\num{1211.36(14640)} & \num{26.67(150)} & \cellcolor{green}\num{0.67(3)} & \cellcolor{green}\num{0.83(2)} & \cellcolor{green}\num{0.45(2)} & \num{0.19(1)} \\ 
	Idriss                & \num{43.88(359)} & \num{1989.69(33127)} & \num{31.66(211)} & \num{0.47(5)} & \cellcolor{gray}\num{0.72(3)} & \num{0.34(4)} & \num{0.22(1)} \\ 
	Aiello                & \cellcolor{gray}\num{44.47(318)} & \cellcolor{gray}\num{2028.23(28522)} & \num{32.67(213)} & \cellcolor{gray}\num{0.45(6)} & \cellcolor{gray}\num{0.72(2)} & \cellcolor{gray}\num{0.32(4)} & \num{0.23(1)} \\ 
	Zhu                   & \num{35.33(245)} & \num{1277.98(17597)} & \cellcolor{green}\num{26.64(190)} & \num{0.66(2)} & \cellcolor{green}\num{0.83(2)} & \cellcolor{green}\num{0.45(2)} & \cellcolor{green}\num{0.18(1)} \\ 
	Mayo                  & \num{40.71(406)} & \num{1739.78(36052)} & \num{29.06(233)} & \num{0.54(6)} & \num{0.76(3)} & \num{0.40(4)} & \num{0.20(2)} \\ 
	Muñoz                 & \num{40.69(186)} & \num{1673.07(14808)} & \cellcolor{gray}\num{32.70(137)} & \num{0.53(6)} & \num{0.82(2)} & \cellcolor{gray}\num{0.32(5)} & \cellcolor{gray}\num{0.26(1)} \\ 
	Khadem                & \num{37.08(225)} & \num{1399.96(16280)} & \num{28.30(152)} & \num{0.62(4)} & \num{0.80(2)} & \num{0.41(3)} & \num{0.21(1)} \\ \midrule 
    \multicolumn{8}{c}{} \\
    \bottomrule
    \end{tabular}}
    \label{tab:performance60}
\end{table*}
\begin{table*}[htp!]
    \centering
        \caption{Results of the ten NN models for each performance metric for $\mathrm{ph}=$\SI{120}{\min}. Green cells in each column highlight the model with the best performance, whereas the grey coloured cells are the worst.}

    \resizebox{14cm}{!}{\begin{tabular}{ | l | c | c | c | c | c | c | c | }
\toprule
    \multicolumn{1}{c}{}   & \multicolumn{7}{c}{Metrics}             \\  \cmidrule{2-8}   
	\multicolumn{1}{c|}{} & RMSE  $\left ( \si{\BGunit} \right )$           & MSE  $\left ( \si{\square\milli\gram\per\square\deci\liter} \right )$                & MAE  $\left ( \si{\BGunit} \right )$            & R$^2$          & CC             & FIT             & MARD          \\ \midrule
	Mirshekarian          & \num{57.43(374)} & \num{3368.68(41488)} & \num{44.08(245)} & \num{0.06(13)} & \num{0.45(6)}  & \num{0.08(6)}   & \cellcolor{green}\num{0.32(2)} \\ 
	Meijner               & \num{57.19(380)} & \num{3343.40(42329)} & \num{44.20(244)} & \num{0.07(13)} & \num{0.45(6)}  & \num{0.08(7)}   & \cellcolor{green}\num{0.33(2)} \\ 
	Gülesir               & \num{55.98(176)} & \num{3172.05(30123)} & \num{44.03(165)} & \num{0.11(11)} & \cellcolor{green}\num{0.48(4)}  & \num{0.08(5)}   & \cellcolor{green}\num{0.32(2)} \\ 
	Sun                   & \num{55.70(298)} & \num{3146.97(32700)} & \cellcolor{green}\num{43.76(181)} & \num{0.13(10)} & \num{0.40(10)} & \cellcolor{green}\num{0.09(5)}   & \cellcolor{green}\num{0.32(2)} \\ 
	Idriss                & \cellcolor{gray}\num{65.79(402)} & \cellcolor{gray}\num{4409.79(53892)} & \num{49.97(278)} & \cellcolor{gray}\num{0.00(14)} & \num{0.33(5)}  & \num{-0.04(6)}  & \num{0.36(2)} \\ 
	Aiello                & \num{63.79(353)} & \num{4131.40(45388)} & \num{48.63(173)} & \num{0.00(10)} & \num{0.29(5)}  & \num{-0.01(5)}  & \num{0.35(2)} \\ 
	Zhu                   & \num{56.41(231)} & \num{3209.40(25056)} & \num{45.63(166)} & \num{0.09(12)} & \num{0.47(4)}  & \num{0.05(6)}   & \num{0.36(2)} \\ 
	Mayo                  & \num{61.11(345)} & \num{3793.74(42721)} & \num{46.82(192)} & \cellcolor{gray}\num{0.00(10)} & \num{0.38(4)}  & \num{0.02(5)}   & \num{0.33(2)} \\ 
	Muñoz                 & \cellcolor{green}\num{54.70(227)} & \cellcolor{green}\num{3017.39(24741)} & \num{45.05(179)} & \cellcolor{green}\num{0.16(8)}  & \num{0.56(5)}  & \num{0.06(5)}   & \num{0.37(2)} \\ 
	Khadem                & \num{64.83(317)} & \num{4253.60(42847)} & \cellcolor{gray}\num{50.56(160)} & \cellcolor{gray}\num{0.00(12)} & \cellcolor{gray}\num{0.20(4)}  & \cellcolor{gray}\num{-0.06(6)}  & \cellcolor{gray}\num{0.38(3)} \\ \midrule 
	\multicolumn{8}{c}{} \\
	\bottomrule
    \end{tabular}}
    \label{tab:performance120}
\end{table*}

On the other hand, in clinical practice, physicians usually plot predictions versus actual values using the Parkes error grid (PEG) \cite{parkes}. This graph has five zones (A to E) to bound prediction accuracy. These zones are set by taking into account the treatment applied for a corresponding BG level. While zone A will always correspond to a correct treatment, zone E corresponds to a hyperglycemia treatment while the patient will suffer from hypoglycemia, or vice versa.

\begin{figure*}[htp]%
\centering
\begin{tabular}{ c c c }
\subfloat[Sun PEG ph = \SI{30}{\min}]{ %
        \includegraphics[width=0.3\linewidth]{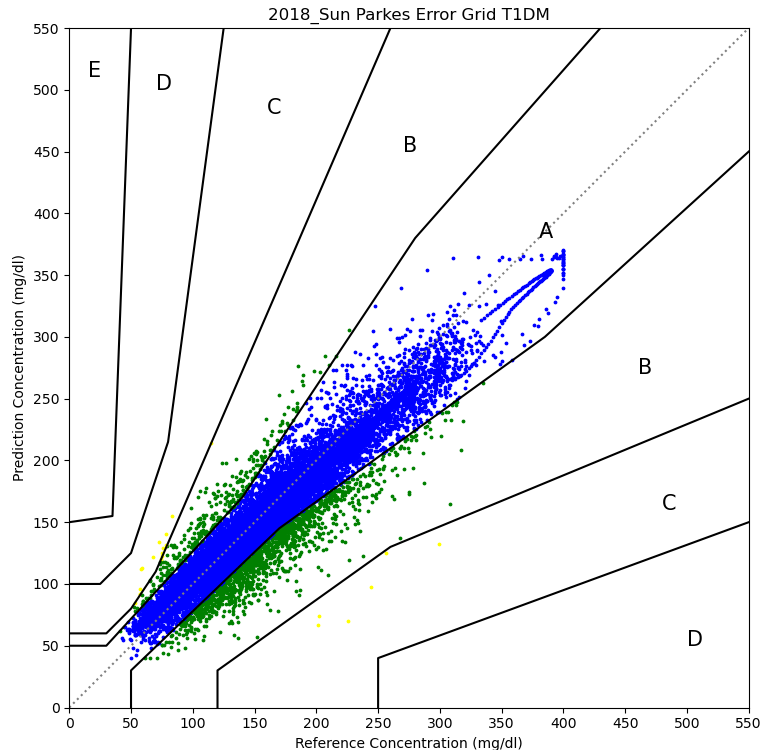}
        \label{fig:SunPEG}
} &
\subfloat[Sun PEG ph = \SI{60}{\min}]{ %
        \includegraphics[width=0.3\linewidth]{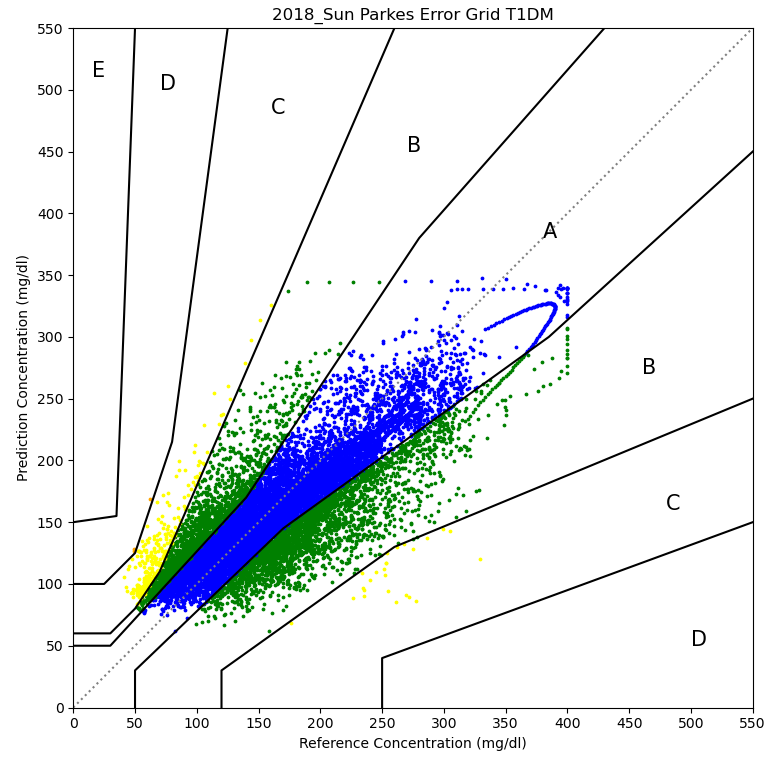}
        \label{fig:SunPEG60}
} &
\subfloat[Sun PEG ph = \SI{120}{\min}]{ %
        \includegraphics[width=0.3\linewidth]{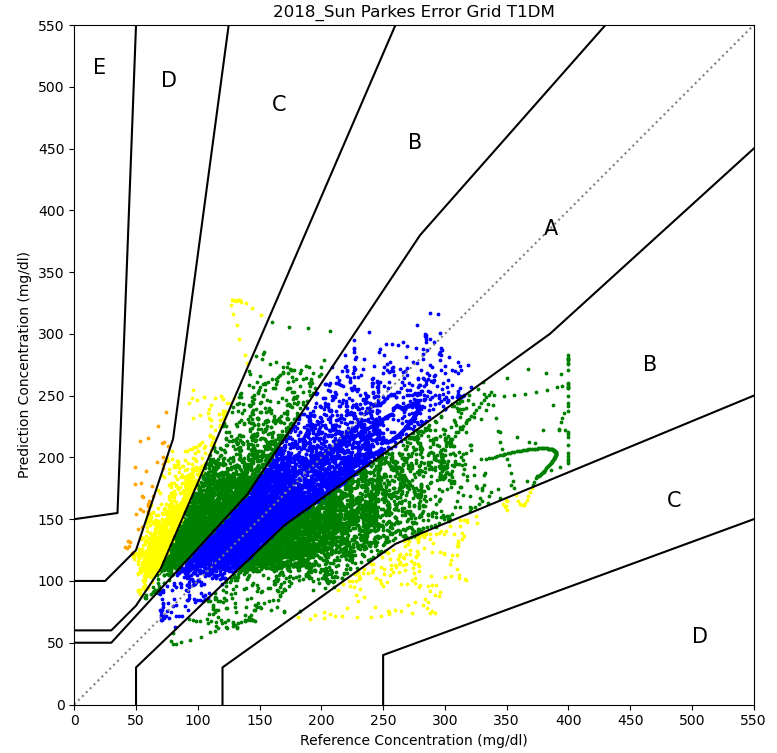}
        \label{fig:SunPEG120}
}
\\
\subfloat[Muñoz PEG ph = \SI{30}{\min}]{ %
        \includegraphics[width=0.3\linewidth]{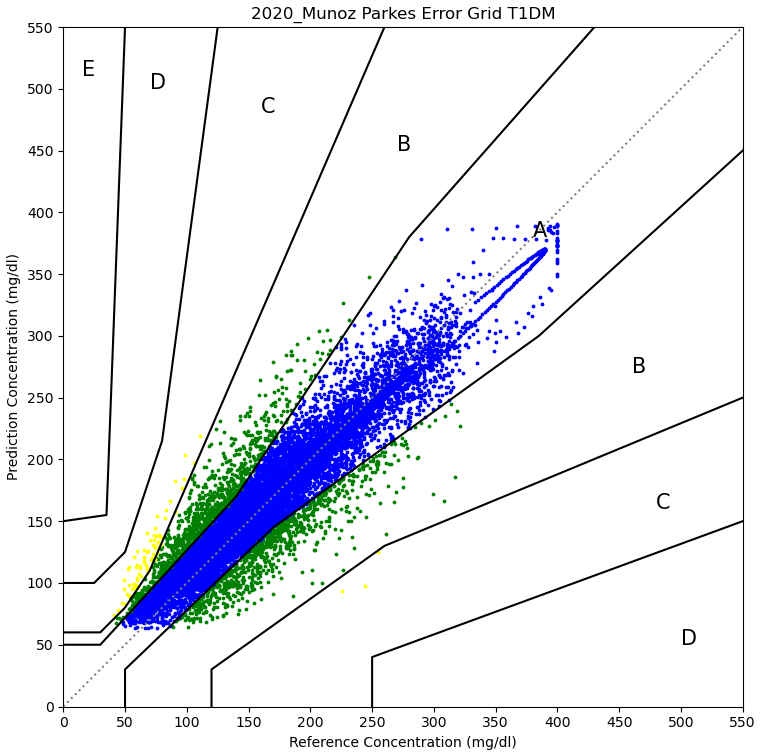}
        \label{fig:MunozPEG}
} &
\subfloat[Muñoz PEG ph = \SI{60}{\min}]{ %
        \includegraphics[width=0.3\linewidth]{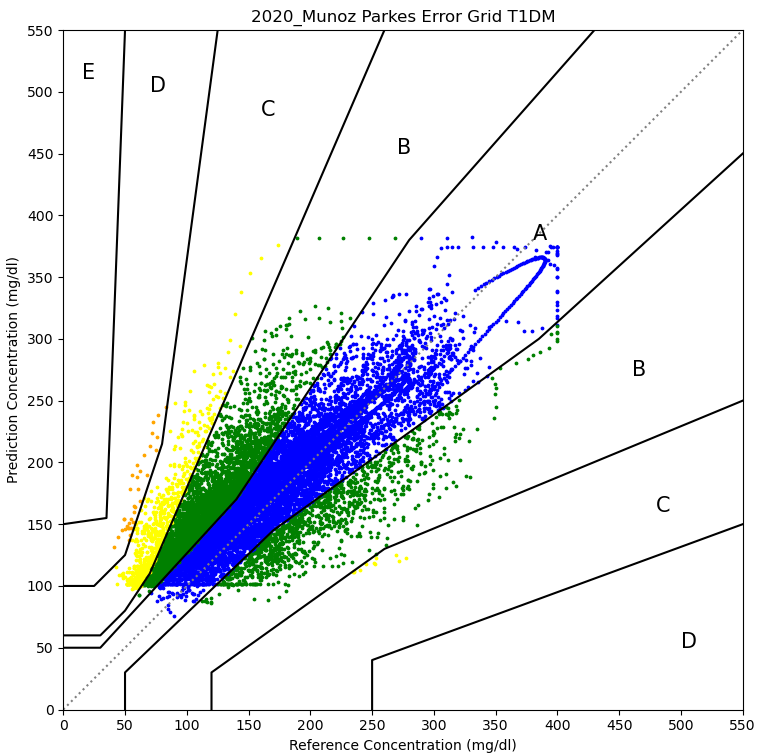}
        \label{fig:MunozPEG60}
} &
\subfloat[Muñoz PEG ph = \SI{120}{\min}]{ %
        \includegraphics[width=0.3\linewidth]{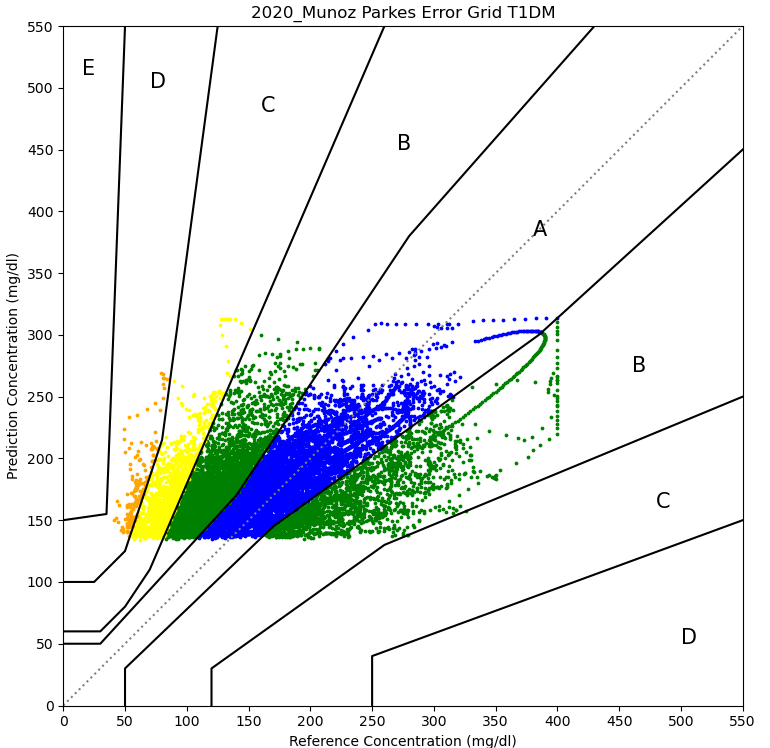}
        \label{fig:MunozPEG120}
}
\end{tabular}
\caption{Parkes Error grid for the models with the highest and lowest number of predictions in A zone.}
\label{fig:grid_analysis}
\end{figure*}

\Cref{fig:grid_analysis} illustrates the PEG for the models with the highest and lowest number of predictions in region A for the three prediction horizons. Hence, \Crefrange{fig:SunPEG}{fig:MunozPEG120} compare (Sun, 2018) versus (Muñoz, 2020). For $ph=\SI{30}{\min}$ the first one has \SI{99.86}{\percent} of predictions between zones A and B versus the second one with \SI{99.52}{\percent}. For $ph=\SI{60}{\min}$, we obtain a range between \SI{97.7}{\percent} and \SI{93.6}{\percent} of points inside regions A and B. Finally, for $ph=\SI{120}{\min}$, in the worst case, \SI{86.2}{\percent} of the points lie in regions A and B, while in the best case \SI{89.5}{\percent} are within them. From this analysis we can conclude that in all the PHs every model predicts well from the clinical point of view even though the differences between the models within a PH are negligible.
In essence, all the metrics show consistent values within their error metrics. However, the confidence intervals overlap, so we cannot conclude which models are better. From the predictors' point of view the models predict well for $\mathrm{ph}=$\SI{30}{\min} and $\mathrm{ph}=$\SI{60}{\min} but for $\mathrm{ph}=$\SI{120}{\min} the models have no accuracy at all. From the clinical point of view the models show no difference between choosing one model or the other. For $\mathrm{ph}=$\SI{30}{\min} and $\mathrm{ph}=$\SI{60}{\min} the models have very good accuracy and, in contrast with the first point of view, we can still use $\mathrm{ph}=$\SI{120}{\min}.

We cannot extract definitive conclusions about NN performance by only using previous metrics since the confidence intervals of most of the metrics in previous tables overlap. This can be explained by the amount of data available in which there are not enough possible scenarios to be learned by the models. To extract these conclusions, we use three comparison methods based on the losses of the predictions, each one applying a different statistical approach, either frequentist or Bayesian. In all three of them, RMSE is the metric of choice to estimate prediction losses. Using these three methods, the models are compared for $\mathrm{ph}=\SI{30}{\min}$, $\mathrm{ph}=\SI{60}{\min}$, and $\mathrm{ph}=\SI{30}\cup\SI{60}{\min}$; the last is a global evaluation of  NN performance in a multi-horizon approach.

\begin{figure*} [htp]
    \centering
    \begin{tabular}{ c c }
    \subfloat[ph = \SI{30}{\min}]{
        \includegraphics[width=0.5\linewidth]{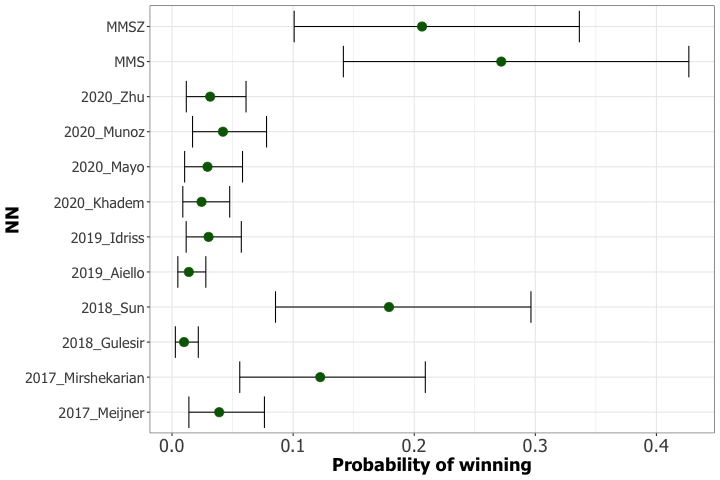}
        \label{fig:Bayesian30}
    } \hfill &
    
    \subfloat[ph = \SI{60}{\min}]{
        \includegraphics[width=0.5\linewidth]{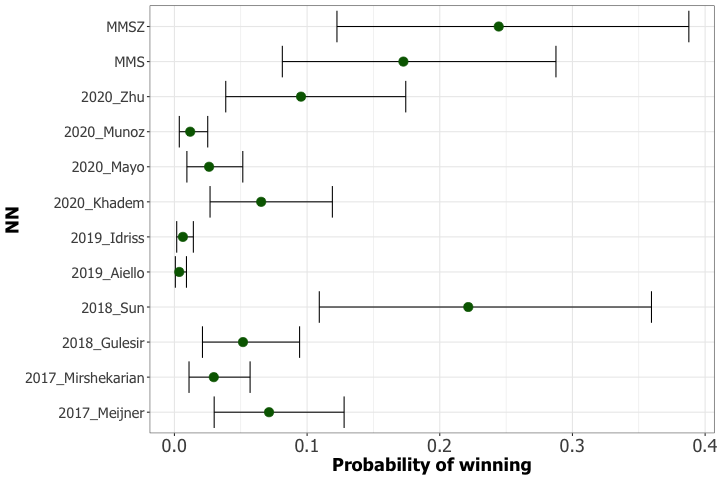}
        \label{fig:Bayesian60}

    } \hfill \\
    \multicolumn{2}{c}{
    \subfloat[Multi-horizon]{
        \includegraphics[width=0.5\linewidth]{images/Bayesian_RMSE30_using_patients.png}
        \label{fig:BayesianMulti}
    } \hfill}
    \end{tabular}
    \caption{Ranking of the best NN using the scmamp method for different prediction horizons.}     
    \label{fig:Bayesian}
\end{figure*}

The scmamp method \cite{Calvo2016, Calvo2018}, is a Bayesian approach based on Plackett-Luce (PL) distribution over rankings to analyse different models regarding multiple problems. The method proposes to use the PL model with a Dirichlet prior to estimating the expected probability of a model of being the best, i.e., the probability of winning. The selection of the best model is based only on its ranking and this method does not take into account the magnitude of the difference between the prediction loss of the different models. We run it twice. First, we compare the ranking of all the NNs in the previous section and, using the results, we create two ensemble methods with the best-ranked models. The two ensemble models are denoted by the initials of their NN models. Hence, MMS stands for the ensemble model that aggregates Mirshekarian, Meijner, and Sun NNs, whereas MMSZ stands for the ensemble consisting of Mirshekarian, Meijner, Sun, and Zhu. In the second run, we compare the models, including the two ensembles. \Crefrange{fig:Bayesian30}{fig:BayesianMulti} show the probability of each model being the best for \SI{30}{\min}, \SI{60}{\min}, and multi-horizon respectively. According to these results, Sun's model is the model with the highest probability of being the best one among the non-ensemble models at all prediction horizons; for example, a probability in the range of \num{0.09} and \num{0.3} for \SI{30}{\min}. At ph=\SI{30}{\min}, Mirshekarian is the second best model clearly separated from the remainder of models whereas at ph=\SI{60}{\min}, the second best model is not so clear when Zhu, Khadem, and Meiner are competing for this rank. Regarding the ensemble models, both of them perform similarly, although MMS, the model with the lowest number of models, has the highest probability of winning in a multi-horizon scenario, with a probability as high as \num{0.48}.

The Model Confidence Set (MCS) \cite{hansenMCS} is a frequentist method whose aim is to determine which models are the best within a collection of models with a given level of confidence, analogous to the confidence interval for a parameter. It consists of a series of tests that repeatedly filter the models in the initial test to finally return the set of those with the lowest losses with confidence level $\alpha$, which we denote as $\mathrm{MCS}^{\alpha}_{ph}$. The tests are run on a sample of the models' predictions, typically using bootstrap replications. Particularly, in this work, we set 1000 bootstrap replications and $ \alpha = 0.05 $, that is, the models with a p-value$>$\num{0.05} are in the confidence set and, with different probabilities, they will return the best predictions. In this case, $\mathrm{MCS}^{0.05} = \left \{ \mathrm{Meijner}, \mathrm{Mayo}, \mathrm{MMS} \right \}$ for all the prediction horizons as well as the multi-horizon analysis.

The Superior Predictive Ability (SPA) \cite{hansenSPA} uses the model's losses as a benchmark, and its null hypothesis is that any model is better than the benchmark. Hence, if the p-value is high there are no models better than the benchmark. This algorithm returns three p-values: lower, consistent and upper. They correspond to different re-centerings of the losses and, normally, the consistent one is the value taken into account \cite{white}. For this algorithm we can say that the best models are Meijner, Sun, Mayo and both ensembles.

Summarizing all the findings, 
\begin{enumerate}

    \item The comparison of the models based only on confidence intervals or the distribution of predictions in the grid's regions is not precise enough to rank the models. Indeed, the difference between the best and worst models is only \SI{3.84}{\BGunit} for RMSE$_{30}$, and \SI{9.99}{\BGunit} RMSE$_{60}$, which although notable for a prediction model, is irrelevant to the physicians' practice. 
    \item At \SI{30}{\min}, the best models are consistently (either using scmamp, MCS, or SPA) the ensemble models, Sun, and Mirshekarian. The ensembles have a higher probability of winning, but their ranges of probability overlap with Sun's range. So, taking into account the complexity of the ensemble models, Sun's model can be a reasonable choice that combines good predictions with lower complexity.
    \item At \SI{60}{\min}, the best models are the ensemble models, Sun, and Zhu. As stated above, attending to information criteria to select the best model with the lowest complexity, Sun seems to be the best option.
    \item None of the NN models provide accurate predictions \SI{120}{\min} ahead of time.
\end{enumerate}

\begin{table}[htp!]
    \centering
        \caption{P-values for the SPA using each model as benchmark for $ph=30,60,30\cup60$. Green cells in each column highlight the model with the best performance.}

    \begin{tabular}{ | l | c | c | c | c | }
\toprule
    \multicolumn{1}{c}{}   & \multicolumn{4}{c}{P-values} \\  \cmidrule{3-5}   
	\multicolumn{2}{c|}{}                   & 30   & 60   & MH    \\ \midrule
	\multirow{3}{*}{Mirshekarian}  & Lower  & 0.18 & 0.00 & 0.01  \\ & Consistent  & .23 & 0.00 & 0.01  \\ & Upper & 0.45 & 0.05 & 0.05  \\ \midrule
	\multirow{3}{*}{Meijner}       & Lower  & 0.44 & 0.54 & 0.54  \\ & Consistent  & \cellcolor{green}0.62 & \cellcolor{green}0.93 & \cellcolor{green}0.84  \\ & Upper & 0.79 & 0.99 & 0.99  \\ \midrule
	\multirow{3}{*}{Gülesir}       & Lower  & 0.00 & 0.00 & 0.00  \\ & Consistent  & 0.00 & 0.00 & 0.00  \\ & Upper & 0.00 & 0.02 & 0.00  \\ \midrule
	\multirow{3}{*}{Sun}           & Lower  & 0.19 & 0.08 & 0.12  \\ & Consistent  & \cellcolor{green}0.21 & \cellcolor{green}0.21 & \cellcolor{green}0.15  \\ & Upper & 0.39 & 0.38 & 0.31  \\ \midrule
	\multirow{3}{*}{Idriss}        & Lower  & 0.01 & 0.00 & 0.00  \\ & Consistent  & 0.01 & 0.01 & 0.00  \\ & Upper & 0.01 & 0.01 & 0.00  \\ \midrule
	\multirow{3}{*}{Aiello}        & Lower  & 0.04 & 0.00 & 0.00  \\ & Consistent  & 0.07 & 0.00 & 0.00  \\ & Upper & 0.09 & 0.00 & 0.00  \\ \midrule
	\multirow{3}{*}{Zhu}           & Lower  & 0.00 & 0.06 & 0.00  \\ & Consistent  & 0.00 & 0.06 & 0.00  \\ & Upper & 0.00 & 0.16 & 0.00  \\ \midrule
	\multirow{3}{*}{Mayo}          & Lower  & 0.43 & 0.12 & 0.17  \\ & Consistent  &\cellcolor{green} 0.53 & \cellcolor{green}0.13 & \cellcolor{green}0.18  \\ & Upper & 0.76 & 0.19 & 0.30  \\ \midrule
	\multirow{3}{*}{Muñoz}         & Lower  & 0.00 & 0.00 & 0.00  \\ & Consistent  & 0.00 & 0.00 & 0.00  \\ & Upper & 0.00 & 0.00 & 0.00  \\ \midrule
	\multirow{3}{*}{Khadem}        & Lower  & 0.00 & 0.06 & 0.00  \\ & Consistent  & 0.00 & 0.15 & 0.00  \\ & Upper & 0.00 & 0.31 & 0.00  \\ \midrule
    \multirow{3}{*}{ensemble-MMS}  & Lower  & 0.73 & 0.19 & 0.37  \\ & Consistent  & \cellcolor{green}0.95 & \cellcolor{green}0.52 & \cellcolor{green}0.63  \\ & Upper & 1.00 & 0.93 & 0.99  \\ \midrule
    \multirow{3}{*}{ensemble-MMSZ} & Lower  & 0.27 & 0.18 & 0.16  \\ & Consistent  & \cellcolor{green}0.33 & \cellcolor{green}0.48 & \cellcolor{green} 0.39  \\ & Upper & 0.59 & 0.83 & 0.82  \\ \midrule
	\multicolumn{5}{c}{} \\
	\bottomrule
    \end{tabular}
    \label{tab:SPA}
\end{table}

\begin{table}[htp!]
    \centering
        \caption{P-values for the MCS for $ph=30,60,30\cup60$ with 1000 bootstrap replications and $ \alpha = 0.05 $ }

    \begin{tabular}{ | l | c | c | c | }
\toprule
    \multicolumn{1}{c}{}   & \multicolumn{1}{c}{} \\  \cmidrule{2-4}   
	\multicolumn{1}{c|}{} & 30   & 60   & MH  \\ \midrule
	Mirshekarian          & 0.02 & 0.00 & 0.00 \\
	Meijner               & \cellcolor{green} 0.36 & \cellcolor{green} 1.00 & \cellcolor{green} 1.00 \\
	Gülesir               & 0.00 & 0.00 & 0.00 \\
	Sun                   & 0.00 & 0.00 & 0.00 \\
	Idriss                & 0.00 & 0.05 & 0.00 \\
	Aiello                & 0.03 & 0.00 & 0.00 \\
	Zhu                   & 0.00 & 0.00 & 0.00 \\
	Mayo                  & \cellcolor{green} 0.87 & \cellcolor{green} 0.32 & \cellcolor{green} 0.44 \\
	Muñoz                 & 0.00 & 0.00 & 0.00 \\
	Khadem                & 0.00 & 0.05 & 0.00 \\ 
	ensemble-MMS          & \cellcolor{green} 1.00 & \cellcolor{green} 0.58 & \cellcolor{green} 0.67 \\
	ensemble-MMSZ         & 0.00 & \cellcolor{green} 0.58 & 0.01 \\\midrule
	\multicolumn{4}{c}{} \\
	\bottomrule
    \end{tabular}
    \label{tab:MCS30}
\end{table}

\section{Conclusions}\label{sec:conclusions}

This article compares ten recently proposed NNs and proposes two ensemble NN-based models for BG prediction. All of them are tested under the same conditions using the most common analysis tools and metrics in the literature. Likewise, all the models are trained and tested using the OhioT1DM Dataset and three different prediction horizons: 30, 60, and 120 minutes. 

When analyzed utilizing these metrics, we find little difference among the models' performance since their metric values are very close and they have overlapping confidence intervals. Indeed, the differences between the best and worst models are not significant from a clinical perspective, with the difference between them as low as \SI{3.84}{\BGunit} for RMSE at the \SI{30}{\min} prediction horizon, and \SI{9.99}{\BGunit} at \SI{60}{\min}.

In contrast, for \SI{120}{\min}, the metrics show that the predictors do not work well, explaining only around \SI{16}{\percent} of glucose variability.

We also analyze the models' performance using the scmamp, MCS, and SPA methods. These analyses consistently show a higher probability of winning for the ensemble-MMS, Sun, and Mirshekarian models for \SI{30}{\min}. For \SI{60}{\min}, the best models are both ensemble models, Sun and Zhu. Finally, for the multi-horizon approach, the best ones are the ensembles, Sun and Mirshekarian. Again, in the three prediction horizons, the results intervals of the best models overlap.

Nevertheless, PEGs show a large number of predictions within zones A and B. Thus, from the clinician's point of view, they can be used.

Therefore, taking into account that the complexity of the models is not a characteristic that improves performance and that there is no differentiation either from the clinical point of view or from the predictive tool point of view, we can state that the best models are Sun, Meijner, and Mirshekarian, as they are the models with the least complexity. Nevertheless, if the NN architecture's complexity is not a critical issue then the ensemble models are the best choice. 

All the models in the best model set share the feature of being a variation of LSTM models: Sun is a Bi-LSTM NN, whereas Mirshekarian is a classical LSTM, Meijner is a customized LSTM, and both ensemble models are a combination of three and four LSTM models respectively. These findings clearly state that this architecture, specifically devised to find a temporal pattern in the input data, is the best option to accomplish future improvement in BG prediction using NN.

In future work, on the one hand, we will implement these models in hardware to obtain a wearable device that can be integrated with a sensor or  insulin pump that meets the power-consumption, durability, and weight parameters of a medical device. On the other hand, we observe a threshold in \Crefrange{fig:SunPEG120}{fig:MunozPEG120} and in every model for ph = \SI{120}{\min}. This also occurs in other LSTM models when there is no sufficient information for the training phase. This leaves open a possible line of research to understand the behavior of neural networks and obtain more efficient training.

\nocite{}

\end{document}